\documentclass[letterpaper]{article} 
\usepackage{aaai2026}  
\usepackage{times}  
\usepackage{helvet}  
\usepackage{courier}  
\usepackage[hyphens]{url}  
\usepackage{graphicx} 
\usepackage{natbib}  
\usepackage{caption} 

\usepackage{array}
\usepackage{physics}
\usepackage{quantikz}
\usepackage[caption=false,font=normalsize,labelfont=sf,textfont=sf]{subfig}
\usepackage{amssymb}
\usepackage{booktabs} 

\usepackage{algorithm}
\usepackage{algorithmic}

\usepackage{newfloat}
\usepackage{listings}
\DeclareCaptionStyle{ruled}{labelfont=normalfont,labelsep=colon,strut=off} 
\floatstyle{ruled}
\newfloat{listing}{tb}{lst}{}
\floatname{listing}{Listing}
\title{Quantum Probabilistic Label Refining: Enhancing Label Quality for \\Robust Image Classification}
\author {
        Fang Qi,
    Lu Peng,
    Zhengming Ding
}
\affiliations {
    Department of Computer Science, Tulane University, New Orleans, Louisiana, USA\\
    fqi2@tulane.edu,
    lpeng3@tulane.edu,
    zding1@tulane.edu
}
    
\nocopyright

\usepackage{bibentry}

\begin{document}

\maketitle

\begin{abstract}

Learning with softmax cross-entropy on one-hot labels often leads to overconfident predictions and poor robustness under noise or perturbations. Label smoothing mitigates this by redistributing some confidence uniformly, but treats all samples equally, ignoring intra-class variability. We propose a hybrid quantum–classical framework that leverages quantum non-determinism to refine data labels into probabilistic ones, offering more nuanced, human-like uncertainty representations than label smoothing or Bayesian approaches. A variational quantum circuit (VQC) encodes inputs into multi-qubit quantum states, using entanglement and superposition to capture subtle feature correlations. Measurement via the Born rule extracts probabilistic soft labels that reflect input-specific uncertainty. These labels are then used to train a classical convolutional neural network (CNN) with soft-target cross-entropy loss. On MNIST and Fashion-MNIST, our method improves robustness—achieving up to 50\% higher accuracy under noise—while maintaining competitive clean-data accuracy. It also enhances model calibration and interpretability, as CNN outputs better reflect quantum-derived uncertainty. This work introduces \textbf{Quantum Probabilistic Label Refining}, bridging quantum measurement and classical deep learning for robust training via refined, correlation-aware labels without architectural changes or adversarial techniques.

\end{abstract}


\section{Introduction}

Deep learning has transformed image classification, yet even top models remain brittle under minor distribution shifts like noise, blur, or rotation \cite{hendrycks2019benchmarking, minderer2021revisiting}. In contrast, human perception remains robust and calibrated under similar conditions \cite{muttenthaler2022human, geirhos2021partial, hermann2020origins, minderer2021revisiting}. A key reason is the use of one-hot labels in training, which treat classification as deterministic, forcing full certainty on a single class, even for ambiguous inputs \cite{zhang2017mixup}. This overconfidence is problematic in high-stakes settings. In autonomous driving, misclassifying a road sign due to occlusion or lighting can be catastrophic. In medical imaging, confident errors on borderline cases may lead to misdiagnoses. In such domains, recognizing and conveying uncertainty is essential \cite{hendrycks2019augmix, ghoshal2021estimating}.

\begin{figure}[h!]
  \centering
  \includegraphics[width=1\linewidth]{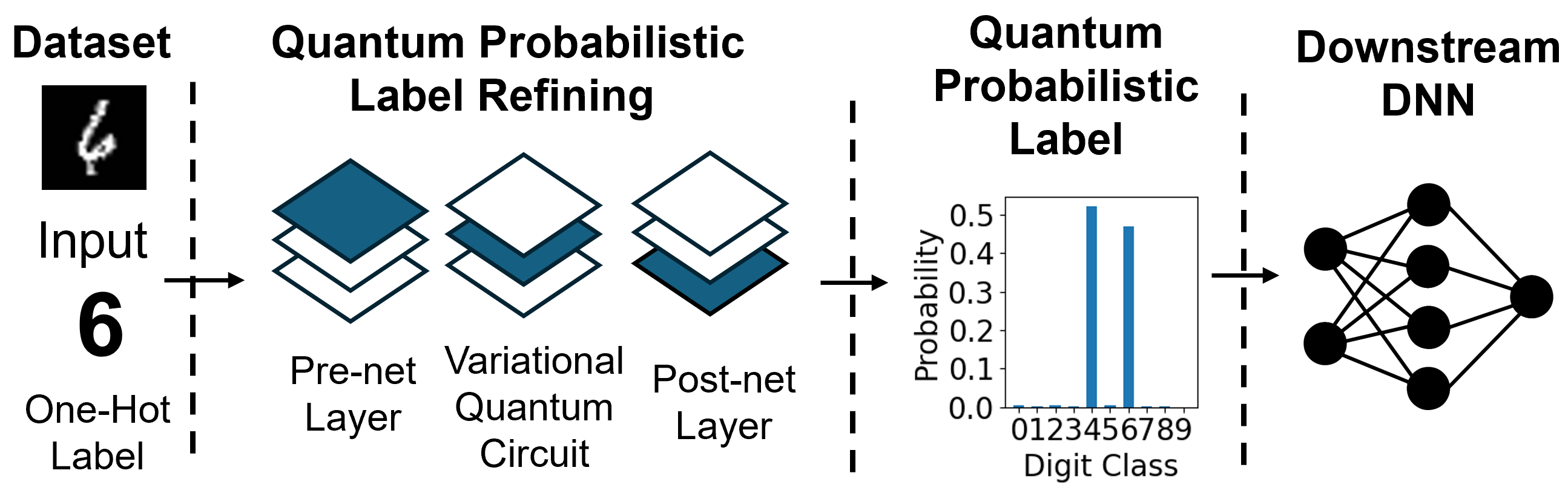}\vspace{-2mm}
  \caption{Conceptual overview of our QPLR approach, where an input digit is encoded into a variational quantum circuit and outputs a \emph{probabilistic label distribution} over the $K=10$ classes. This soft label supervises a downstream DNN to capture more realistic crowd-source knowledge.}\vspace{-3mm}
  \label{fig:overview}
\end{figure}

Label smoothing is a common solution that redistributes a fraction of confidence from the true class to others to reduce overconfidence \cite{szegedy2016rethinking}. However, it applies the same softened distribution to all samples, ignoring differences in ambiguity among inputs. In reality, class overlaps and uncertain samples are poorly represented in standard datasets. What's missing is a data-dependent method to reflect sample-specific uncertainty—something static label smoothing can't achieve.

To address this, we turn to quantum non-determinism as a natural source of uncertainty. Quantum systems yield probabilistic outputs upon measurement—a principle formalized by the Born rule \cite{hall2013quantum}, which assigns outcome probabilities based on the squared amplitude of quantum states \cite{nielsen2010quantum}. By encoding images into a variational quantum circuit (VQC), we harness quantum superposition and entanglement to produce rich, sample-specific distributions over class labels \cite{huang2021power, mcardle2020quantum, cong2019quantum}. Specifically, entanglement encodes input-feature correlations, and superposition represents diverse configurations. Quantum measurement then extracts input-dependent probabilistic labels that reflect these correlations, offering a more expressive alternative to classical heuristics.

We introduce Quantum Probabilistic Label Refining (QPLR), a novel framework that uses a VQC to generate soft labels from quantum measurement distributions, shown in Fig. \ref{fig:overview}. These are then used to train a classical convolutional neural network (CNN) via soft-target cross-entropy loss. Unlike one-hot or uniform-smoothed labels, these targets reflect input-specific ambiguity and class overlap. To our knowledge, this is the first work to leverage quantum measurement distributions as soft labels for classical deep learning. Our contributions are summarized as follows:
\begin{itemize}
\item \textbf{QPLR framework:} We propose QPLR, a hybrid method using quantum measurement from VQCs to generate refined, sample-specific probabilistic labels for training classical networks.

\item \textbf{Improved robustness:} CNNs trained with QPLR show up to 50\% better accuracy under image corruptions on MNIST and Fashion-MNIST, without degrading clean-data performance.

\item \textbf{Better calibration:} QPLR enhances uncertainty estimation and interpretability, critical for safety-sensitive applications like autonomous driving and medical AI.
\end{itemize}

\section{Related Work}
\label{sec:related}

\vspace{1mm}\noindent\textbf{Quantum Deep Learning}. Variational Quantum Circuits (VQCs) have emerged as a novel approach for realizing near-term quantum advantage in supervised learning tasks \cite{cerezo2021variational, blance2021quantum, maheshwari2021variational, kapoor2016quantum, khairy2020learning}. 
Compared to traditional quantum neural networks (QNNs) \cite{abbas2021power, henderson2020quanvolutional, cong2019quantum} that rely on mid-circuit measurements or deep architectures, VQCs prioritize resource efficiency on NISQ devices by using shallow parameterized circuits without intermediate state collapse, thereby mitigating decoherence and measurement noise \cite{cerezo2021variational, benedetti2019parameterized}.
Early works successfully demonstrated VQCs on small-scale benchmarks such as MNIST, achieving feasibility on simulators and noisy hardware \cite{havlivcek2019supervised, caro2022generalization, chalumuri2021hybrid,  zhang2023statistical, schuld2020circuit, mcardle2019variational}. However, these efforts have focused on using the quantum model itself as the final classifier—i.e., collapsing the measurement outcome into a single label or expectation value. In contrast, our work exploits the \emph{full probabilistic output} of the VQC (i.e., the distribution over measurement outcomes) as a \emph{teaching signal} for a classical network.

\vspace{1mm}\noindent\textbf{Soft Labels and Probabilistic Supervision}. In classical deep learning, \emph{soft labels} or probability targets are well known to improve generalization and calibration.  The seminal work of Hinton et al.\ introduced \emph{knowledge distillation}, where a large teacher network’s predictive distribution (soft targets) guides a smaller student model, yielding better performance than hard labels alone \cite{hinton2015distilling}.  Label smoothing extends this idea by blending the one-hot target with a uniform distribution over classes, thus preventing overconfidence and improving calibration \cite{szegedy2016rethinking, pereyra2017regularizing}.  More recent studies have shown that soft targets encode inter-class similarity information that hard labels lack, making models more robust to input perturbations \cite{muller2019does, pereyra2017regularizing}.  Our approach can be viewed as a form of distillation where the \emph{teacher is a quantum model}—a fundamentally different source of uncertainty stemming from quantum mechanics rather than a larger neural network.

\vspace{1mm}\noindent\textbf{Robust Image Classification}. Robustness to input noise and geometric distortions remains a central challenge in image classification.  Classical defenses often employ data augmentation—such as additive noise, small‐angle rotations, and elastic distortions—to improve generalization on digit and large‐scale benchmarks \cite{simard2003best, krizhevsky2017imagenet}.  Systematic evaluations, for example via the ImageNet‐C benchmark, have revealed that modern convolutional networks degrade sharply under even modest common corruptions (e.g., Gaussian noise, blur, brightness changes) \cite{hendrycks2019benchmarking}.  To address adversarial threats, adversarial training methods minimize worst‐case loss within an $\ell_\infty$ perturbation ball \cite{madry2017towards}, while deep ensembles of independently trained models can further enhance both robustness and uncertainty estimation at considerable computational cost \cite{lakshminarayanan2017simple}.  In contrast, our approach introduces a complementary \emph{quantum‐inspired regularization}: by supervising a classical CNN with probabilistic labels produced by a variational quantum circuit, the network inherently learns full output distributions—akin to an implicit ensemble—thereby achieving improved resilience to both common corruptions and geometric transformations without altering model architecture or requiring adversarial examples.

\begin{figure*}[!t]
\centering
\subfloat[]{\includegraphics[width=0.2\textwidth, height=3.6cm]{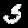} \label{fig:quantumexampleA}}
\hfil
\subfloat[]{\includegraphics[width=0.45\textwidth, height=3.6cm]{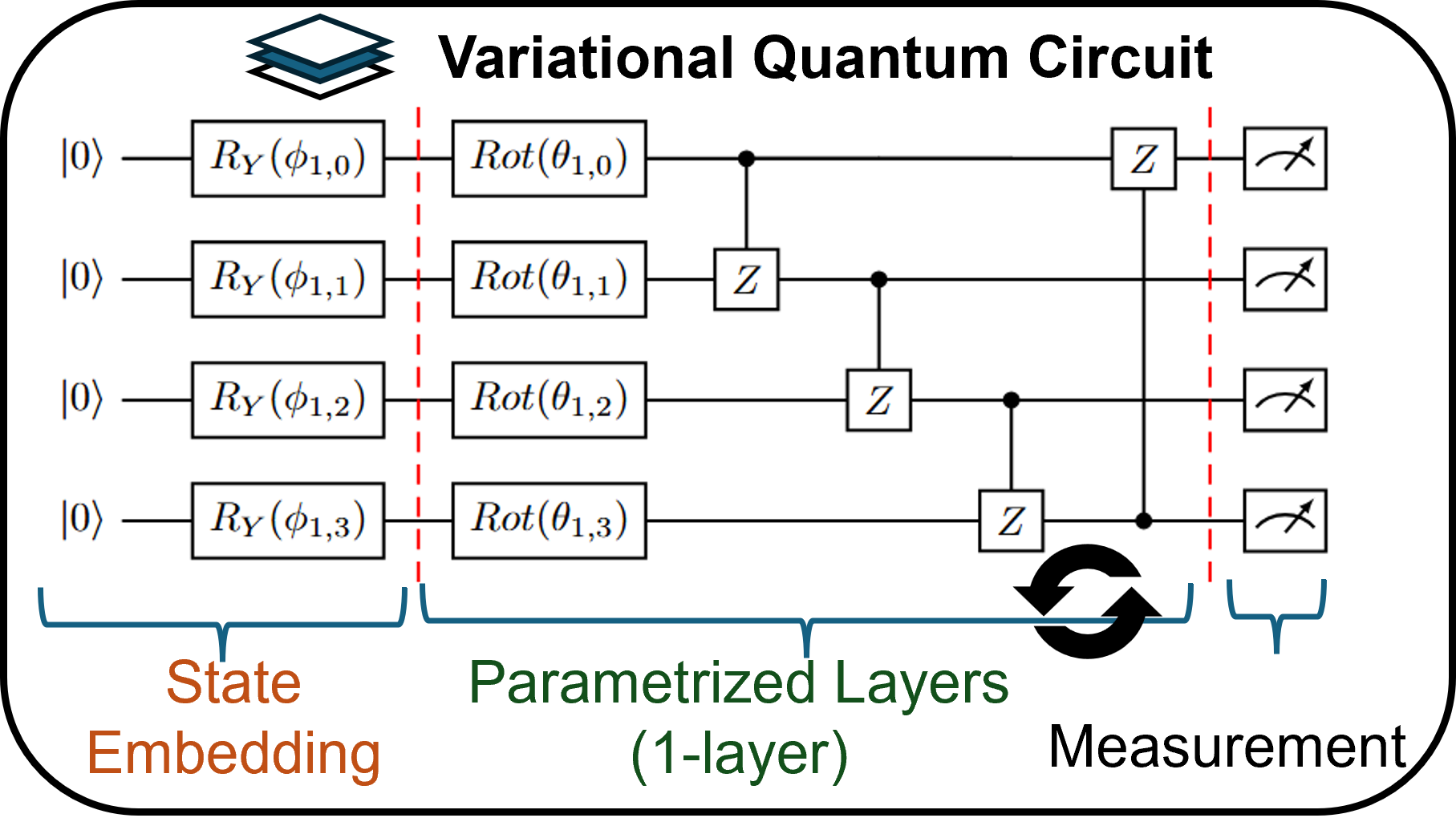} 
\label{fig:quantumexampleB}}
\hfil
\subfloat[]{\includegraphics[width=0.3\textwidth, height=3.6cm]{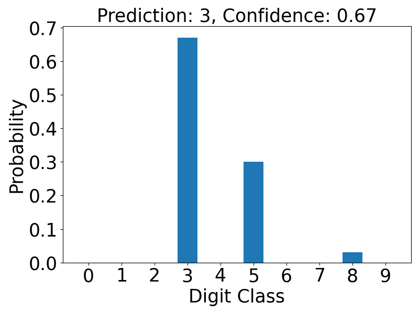} \label{fig:quantumexampleC}}\vspace{-1mm}
\caption{ An illustrative example of generating a soft Quantum labeling. 
(a) An ambiguous MNIST digit “5”.
(b) A simple 4-qubit variational quantum circuit composed of state embedding, one layer of parametrized circuit, and measurement. 
(c) The soft label contains distributions that provide both 3 and 5 as the predictions with higher probability than other digits. }
\label{fig:quantumexample}\vspace{-2mm}
\end{figure*}

\vspace{1mm}\noindent\textbf{Quantum–Classical Hybrid Learning}. Several recent efforts have integrated quantum circuits into classical neural pipelines, for instance, as trainable feature maps \cite{schuld2020circuit, abbas2021power} or hybrid layers \cite{grant2018hierarchical, henderson2020quanvolutional, henderson2021methods}. These hybrid models typically incorporate quantum subroutines directly into the forward pass of the classifier \cite{bokhan2022multiclass, ahmed2023multiclass, ren2022experimental, ding2024quantum}. In contrast, our hybridization occurs at the \emph{label level}: we treat the quantum circuit as a \emph{black-box label generator}.  Once soft labels are generated—potentially offline on quantum hardware or high‐performance simulators—the main training remains entirely classical and scalable on future quantum devices. To our knowledge, using quantum measurement distributions to supervise classical deep networks and exploit Born‐rule randomness for improved learning is novel.

\section{Motivation}
While prior hybrid quantum–classical work has explored variational quantum circuits (VQCs) for classification, few have investigated their potential for \emph{label refinement}. We argue that quantum mechanisms—\textbf{entanglement}, \textbf{superposition}, and \textbf{measurement}—make VQCs uniquely suited to generate soft, probabilistic labels that reflect ambiguity and inter-class relationships in real-world data.

\textbf{Entanglement} allows quantum circuits to encode rich correlations between qubits representing different image regions or features. For $n$ qubits, there are $2^n - n - 1$ possible non-trivial entangled subsets. At $n = 10$, this yields 1013 distinct correlation structures, including pairwise, triplet, and global entanglement. Such expressiveness enables modeling of complex dependencies that classical models struggle to represent.

\textbf{Superposition}, achieved through quantum gates (e.g., $R_Y$, $R_Z$, $U_3$), encodes classical inputs into a $2^n$-dimensional Hilbert space. This permits simultaneous representation of all basis states. Combined with entanglement, it enables highly expressive quantum states, allowing the circuit to explore varied label interpretations and capture fine-grained intra-class variation.

\textbf{Measurement}, governed by the Born rule, collapses the quantum state into a class probability distribution. Unlike deterministic hard labels or fixed label-smoothing heuristics, these measurement-induced distributions vary by input and reflect the uncertainty encoded by the quantum state. They provide a principled way to produce \emph{probabilistic soft labels} that convey richer, sample-specific supervision.

Recent theoretical works validate this advantage. Huang et al.~\cite{huang2021power} demonstrate that random quantum circuits can produce kernel functions beyond the reach of classical approximations. Abbas et al.~\cite{abbas2021power} show that quantum models can linearly separate classes in Hilbert spaces that are otherwise inaccessible to classical networks. These insights suggest that quantum circuits offer powerful data representations with exponentially fewer parameters.

\section{Proposed Method}

\subsection{Quantum Probabilistic Label Refining}
\label{ssec:quantum}

Our first step is to train a \emph{Variational Quantum Circuit} (VQC) on the MNIST training set and then use it as a \emph{quantum labeler}.  Concretely, let $x\in\mathbb{R}^{28\times28}$ be an input digit reshaped to a vector in $\mathbb{R}^{d}$ (with $d=28^2$).  We allocate $n$ qubits so that $2^n\ge K=10$, and define two possible encodings:

\noindent\textbf{Angle encoding:} A classical feed-forward block maps $x$ to an $n$-dimensional rotation vector $\boldsymbol\phi(x)\in[0,\pi]^n$.  We then prepare
  \[
    \ket{\psi(x)} \;=\; \bigotimes_{i=1}^n R_Y\bigl(\phi_i(x)\bigr)\,\ket{0}^{\otimes n},
  \]
  where $R_Y(\phi)$ is a Pauli-$Y$ rotation.  
  
\noindent\textbf{Amplitude encoding:} We first flatten and normalize $x$ to a $2^n$-dimensional state vector $v(x)$, then embed via
  \[
    \ket{\psi(x)} \;=\; \sum_{i=0}^{2^n-1} v_i(x)\,\ket{i}.
  \]

On $\ket{\psi(x)}$ we apply a \emph{parameterized circuit} of $L$ layers, each consisting of
\[
  \underbrace{\bigotimes_{i=1}^n \mathrm{Rot}(\theta_{l,i})}_{\text{single-qubit rotations}}
  \quad\text{followed by}\quad
  \underbrace{\prod_{(i,j)\in\mathcal{E}} \mathrm{CZ}_{i,j}}_{\text{controlled-Z entanglers}},
\]
where the graph $\mathcal{E}$ encodes either a \emph{linear}, \emph{ring}, or \emph{full} entanglement pattern. The resulting state
\[
  \ket{\psi(x;\boldsymbol\theta)} \;=\; U(\boldsymbol\theta)\,V(x)\,\ket{0}^{\otimes n}
\]
depends on both the input‐encoding unitary $V(x)$ and the trainable unitary $U(\boldsymbol\theta)$.

Finally, we perform $M$ repeated projective measurements in the computational basis $\{\ket{y}\}_{y=0}^{2^n-1}$, yielding outcome $y$ with Born‐rule probability
\begin{equation}
  P(y\mid x) \;=\; \bigl|\bra{y}\,\psi(x;\boldsymbol\theta)\bigr|^2.
  \label{eq:born}
\end{equation}
We then truncate or renormalize these probabilities to the $K=10$ digit classes.  In practice, we sample $M=1{,}000$ shots with PennyLane’s simulator (or hardware) to estimate the distribution $\{P(y\mid x)\}$ without any manual calibration.

Because quantum superposition allows $\ket{\psi(x)}$ to overlap multiple basis states, the VQC naturally produces \emph{non‐deterministic} labels.  For a confusing training example (e.g.\ one labeled “5” that visually resembles a “3”), the estimated distribution might be
\[
  P(5\mid x)\approx0.67,\quad P(3\mid x)\approx0.31,\quad P(y\neq3,5)\approx0.02,
\]
reflecting genuine ambiguity, with an example shown as Fig.~\ref{fig:quantumexample}.  In contrast, a classical softmax‐based CNN forced to one‐hot targets would likely assign $\approx99\%$ to a single class.  Moreover, the smooth manifold of quantum states in Hilbert space ensures that small perturbations of $x$ yield gradual changes in $\ket{\psi(x)}$, so similar inputs have similar label distributions.

We train the VQC’s parameters $\boldsymbol\theta$ via gradient‐based optimization (adjoint differentiation on GPU, parameter‐shift otherwise) to minimize the cross‐entropy
\[
  \mathcal{L}_{\mathrm{Q}}(\boldsymbol\theta)
  = -\sum_{i=1}^N \sum_{k=0}^{K-1} \,y^{(i)}_k\,\log P(k\mid x^{(i)};\boldsymbol\theta),
\]
where $y^{(i)}_k\in\{0,1\}$ are the one‐hot true labels.  Although the VQC’s classification accuracy ($\sim96\%$) may lag that of a CNN, our goal is \emph{not} to beat classical performance but to extract a \emph{probabilistic} supervision signal for downstream training.

\subsection{Human and Foundation AI Assessment}

\begin{figure*}[t]
  \centering
  \includegraphics[width=\linewidth]{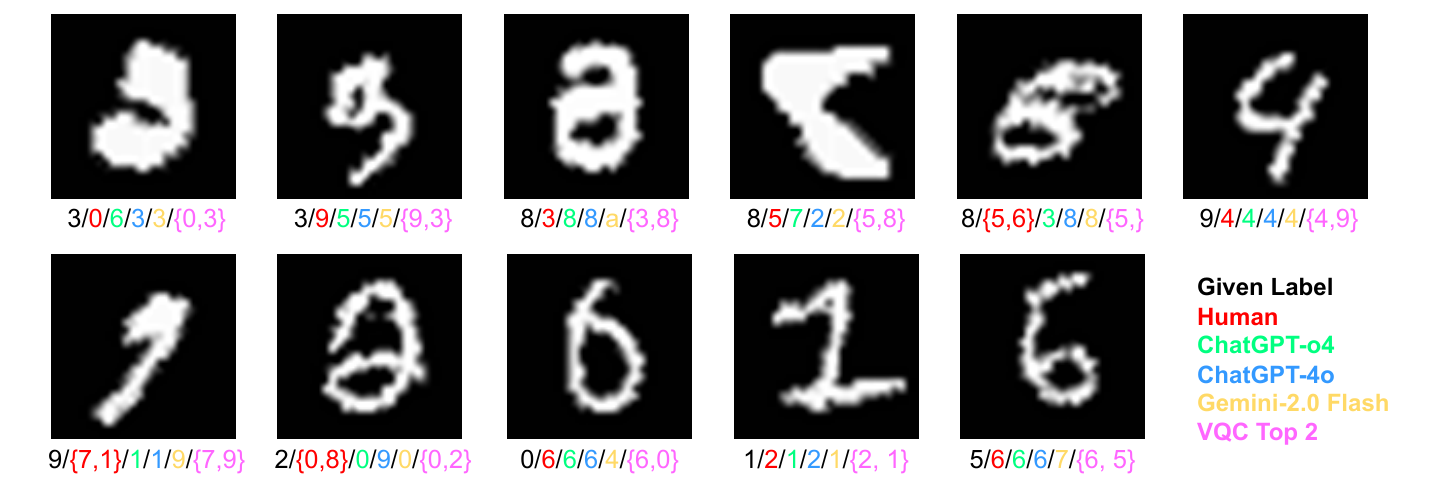}\vspace{-0mm}
  \vspace{-6mm}\caption{Comparison of human, foundation AI model, and quantum probabilistic predictions on ambiguous MNIST train examples. For each digit image, we report the digit recognition assigned by human annotators, ChatGPT-4, ChatGPT-4o, and Gemini 2.0 Flash, and the top 2 classes within the VQC's distribution. These results illustrate differing perceptions of uncertainty across models and humans.[Best viewed in color.]}\vspace{-2mm}
  \label{fig:fig3_assessment}
\end{figure*}

To evaluate the interpretability and reliability of our quantum-derived soft labels, we conducted a two-pronged assessment involving both expert human annotators and large foundation AI models. We selected ambiguous MNIST samples from the training dataset whose quantum-generated label distributions exhibited high entropy (see Figure \ref{fig:fig3_assessment}).

\paragraph{Human Assessment}

The authors independently reviewed each grayscale digit image and assessed its identity based on visual inspection. Using a majority voting scheme, the most likely digit class was selected. In instances where an image exhibited visual ambiguity—appearing similar to more than one digit—multiple plausible classes were recorded. The first digit listed represents the class deemed most likely, while secondary options reflect residual uncertainty. This approach provides a human-centric benchmark for evaluating label ambiguity and comparing against both quantum-derived distributions and AI model predictions.

\paragraph{Foundation Model Assessment}

We queried three state-of-the-art AI models—ChatGPT-o4, ChatGPT-4o, and Gemini-2.0 Flash—using the prompt: \emph{``Please recognize the image as one of the 10 handwritten digit classes.''}

As shown in Fig.\ref{fig:fig3_assessment}, for genuinely ambiguous inputs, the ground-truth label and conventional model outputs sometimes diverge from human perception, whereas our quantum-derived distributions faithfully capture that uncertainty. The VQC assigns its highest probability to the class that humans favor, while still allocating a nonzero mass to the dataset’s annotated label. By contrast, foundation models like Gemini occasionally ignore prompt constraints—for instance, predicting a non‐digit category "a" for an image labeled "2"—likely reflecting internal pattern-recognition biases. In another case, an image annotated as "8" but judged by experts as a "5" or "6" was assigned $>99.9\%$  probability on "5" by the VQC, illustrating its close alignment with human judgment even when it conflicts with the original annotation. These observations demonstrate that even state-of-the-art AI systems can struggle with edge-case ambiguity, highlighting the benefit of probabilistic labeling grounded in quantum nondeterminism.

\subsection{Deep Image Classification}

In image classification tasks, a deep neural network (DNN) typically begins by receiving an input image $\mathbf{x}_i \in \mathbb{R}^{H \times W \times C}$, where $H$, $W$, and $C$ denote the image height, width, and number of channels, respectively. A neural network backbone, such as a convolutional neural network (CNN), processes this input to extract high-level feature representations. These features are then passed to a classifier head—often a fully connected layer followed by a softmax function—to produce a probability distribution over $K$ classes:
\begin{equation}
\mathbf{p}_i = \mathcal{F}_\theta(\mathbf{x}_i) \in [0, 1]^K,
\end{equation}
where $\mathcal{F}_\theta$ denotes the DNN with parameters $\theta$, and $\mathbf{p}_i$ represents the predicted class probabilities for the $i$-th sample.

The classifier is typically trained using the cross-entropy (CE) loss on one-hot labeled data. Let $\mathbf{y}_i \in \{0, 1\}^K$ be the one-hot ground-truth label for the $i$-th input. The CE loss is defined as:
\begin{equation}
\mathcal{L}_\text{ce} = -\frac{1}{N} \sum_{i=1}^{N} \sum_{j=1}^{K} y_{ij} \log(p_{ij})= -\frac{1}{N} \sum_{i=1}^{N} \log(p_{ik}),
\end{equation}
where $p_{ik}$ is the predicted probability for the correct class $k$ of sample $i$. 

However, standard supervised learning typically assumes access to deterministic, one-hot labels \( \mathbf{y}_{\text{one-hot}} \), which encode full confidence in a single class. Training models with softmax cross-entropy loss on such labels often results in overconfident predictions and reduced robustness to input noise or distributional shifts.

The standard softmax cross-entropy loss penalizes deviations from the single "correct" class and ignores the ambiguity inherent in inputs that may resemble multiple classes (e.g., a digit that could reasonably be a ``3'' or an ``8''), meaning one-hot targets can lead to models that are overly confident, often resulting in poor generalization. To mitigate this, \emph{label smoothing} \cite{szegedy2016rethinking} replaces the hard 0/1 targets with smoothed labels, assigning a small portion of probability mass to incorrect classes. Specifically, for a given smoothing factor $\epsilon \in [0, 1]$, the smoothed label becomes:
\begin{equation}
\tilde{y}_{ij} = 
\left\{
\begin{array}{ll}
1 - \epsilon, & \text{if~} j = k \\[4pt]
\displaystyle\frac{\epsilon}{K - 1}, & \text{otherwise}
\end{array}
\right.
\end{equation}

This technique acts as a form of regularization, reducing overfitting and encouraging the model to produce more calibrated probability estimates.

\subsubsection{Probabilistic Labeling via Quantum Non-Determinism}

In real-world classification scenarios, a single input \( x \) may correspond to multiple plausible outcomes \( y \in \{1, \dots, K\} \). To better reflect this uncertainty, we propose using \textit{probabilistic labels} \( \mathbf{y}^{\text{prob}} \in \Delta^{K-1} \), where each label is a probability distribution over the class space. While label smoothing uniformly adjusts all labels by redistributing a fixed amount of confidence, it fails to account for sample-specific ambiguity and treats all inputs identically.

To address this, we introduce a principled mechanism to generate probabilistic labels using \textbf{quantum non-determinism}. Specifically, we employ a variational quantum circuit (VQC) to encode each input \( x \) into a quantum state \( \ket{\psi(x)} \), and extract a class distribution by measuring the state and applying the Born rule \cite{hall2013quantum}. This produces a quantum-derived soft label:
\[
\mathbf{y}^{\text{quantum}} = [p_1^{\text{q}}, \dots, p_K^{\text{q}}],
\]
where each \( p_k^{\text{q}} \) represents the probability of observing class \( k \) given the quantum encoding of \( x \).

We then train a classical neural network using the following soft-label cross-entropy loss:
\begin{equation}
\mathcal{L}'_\text{ce}  = -\sum_{k=1}^{K} y_k^{\text{quantum}} \log p_k,
\end{equation}
where \( \mathbf{p} \) is the predicted distribution from the model. This approach encourages the network to distribute its confidence across plausible classes, leading to improved robustness under noise, better calibration, and more human-like uncertainty estimation.

Such probabilistic modeling is particularly valuable in high-stakes applications. In {autonomous driving}, visual inputs may be ambiguous due to weather, occlusion, or sensor noise, and committing to a single class (e.g., traffic sign type or pedestrian intent) can be dangerous. A probabilistic prediction enables safer decision-making by allowing the system to hedge its actions based on uncertainty. Similarly, in the medical domain, diagnostic inputs such as X-rays or MRIs often admit multiple interpretations. Providing a probability distribution over potential conditions allows clinicians to weigh differential diagnoses more effectively and improves trust in AI-assisted decisions.

\section{Experiments}
\subsection{Experimental Setting}

\subsubsection{Datasets}
We evaluate our methods on two widely used image classification benchmarks.

The MNIST dataset contains 70,000 grayscale images of handwritten digits (0–9), each sized 28×28 pixels. These images are centered and size-normalized, with 60,000 for training and 10,000 for testing. MNIST has long served as a foundational benchmark for assessing model architectures, training strategies, and optimization techniques.

Fashion-MNIST shares the same format and train/test split as MNIST but features 10 categories of clothing items (e.g., shirts, trousers, sneakers). The dataset is significantly more challenging due to its greater visual complexity and intra-class variation, offering a more realistic benchmark for evaluating generalization to natural image textures and shapes.

To better assess model robustness and generalization, we create two challenging variants of these datasets: (a) \textbf{Noisy Variants}: We apply additive Gaussian noise with zero mean and a standard deviation to simulate sensor imperfections or real-world degradation. (b) \textbf{Rotated Variants}: We rotate each image by a fixed angle to evaluate the model's resilience to geometric transformations that often occur in practical settings. 

\subsubsection{Implementation Details}

\paragraph{Quantum Implementation}  
Our variational quantum circuits (VQCs) are defined and executed in \texttt{PennyLane} v0.25 \cite{bergholm2018pennylane}, with a lightweight PyTorch–based pre- and post-network wrapped around each quantum kernel. All experiments ran on the LONI HPC, where each node provides two NVIDIA A100 GPUs and 32 CPU cores. For both MNIST and Fashion-MNIST, each image $\mathbf{x}_i\in\mathbb{R}^{28\times28}$ is flattened to a 784-dimensional vector and passed through a two-layer pre-network (128 units per layer, ReLU activations) that produces $n=10$ rotation angles for \emph{angle encoding} into a 10-qubit register. On MNIST, we stack $L=3$ variational layers—each comprising single-qubit $R_Y$ rotations followed by ring entanglement via Controlled Z gates—and on Fashion-MNIST we use $L=2$. Circuits are measured with $M=1{,}000$ shots, and the resulting counts are renormalized by a two-layer post-network (128→10) to yield a probability distribution over the ten different classes.  

Parameters $\boldsymbol\theta$ are trained with Adam (initial Learning Rate as 0.001, decayed ten-fold after epoch 3) for five epochs on MNIST (batch size 64), reaching 97\% accuracy in approximately 13 hours, and for 25 epochs on Fashion-MNIST, achieving 88.93\% accuracy in around 54 hours. 

\paragraph{Networks Implementation}  
We implement a LeNet-style convolutional neural network for digit classification on MNIST and its variants. The model consists of two convolutional layers: the first applies 6 filters of size 5×5, followed by ReLU activation and 2×2 max pooling, while the second applies 16 filters of size 5×5, again followed by ReLU and 2×2 max pooling. The resulting feature maps are flattened and passed through two fully connected layers with 120 and 84 units, respectively, before reaching the final output layer with 10 units, corresponding to the digit classes. The model is trained using the Adam optimizer with a learning rate of 0.0001 and standard cross-entropy loss, optionally with label smoothing. All experiments are conducted on a machine equipped with two NVIDIA Ada 6000 GPUs, and end-to-end trainings are completed in roughly 10 minutes.

\begin{table}[t] 
\small
\centering
\caption{Performance comparison of methods under varying noise levels.}\label{tb1}\vspace{-2mm}
\centering
\begin{tabular}{|c|c|c|c|c|c|}
\hline
\textbf{Method} & 
\shortstack{\textbf{Std=}\\\textbf{0.1}} & 
\shortstack{\textbf{Std=}\\\textbf{0.2}} & 
\shortstack{\textbf{Std=}\\\textbf{0.3}} & 
\shortstack{\textbf{Std=}\\\textbf{0.4}} & 
\shortstack{\textbf{Std=}\\\textbf{0.5}} \\
\hline
M1 & 99.14\% & 98.12\% & 85.50\% & 62.59\% & 37.68\% \\
M2 & 99.30\% & 98.60\% & 73.51\% & 51.01\% & 23.32\% \\
M3 & 97.87\% & 97.35\% & 95.76\% & 81.34\% & 70.13\% \\
M4 & 97.95\% & 97.82\% & 96.05\% & 80.66\% & 52.82\% \\
\hline
\end{tabular}

\end{table}

We fixed random seeds for NumPy, PyTorch, and PennyLane to ensure deterministic behavior across runs. Code and hyperparameter details are available in our supplementary repository.

\subsection{Comparison Results}

We compare different versions of our LeNet-style convolutional neural network (CNN) trained on MNIST/Fashion-MNIST and their variants, exploring the impact of various techniques on model performance. \textbf{1)} M1 represents the baseline model, where the LeNet is trained with the Softmax activation function for classification. \textbf{2)} M2 enhances M1 by incorporating Label Smoothing \cite{muller2019does}, a regularization technique that slightly adjusts the target labels to improve generalization and reduce overfitting. \textbf{3)} M3 further refines the model by using a Probabilistic Label derived from Quantum AI, which introduces probabilistic distributions over the labels rather than hard labels by using All Training Samples. \textbf{4)} M4 is a variant of M3, with a focus on using only the High-Confident Training Samples (54,887 out of 60k, the confidence threshold is 0.9), allowing the model to concentrate on examples with higher certainty, which might reduce noise and improve training efficiency. 

\begin{figure*}[h!]
  \centering
  \includegraphics[width=1\linewidth]{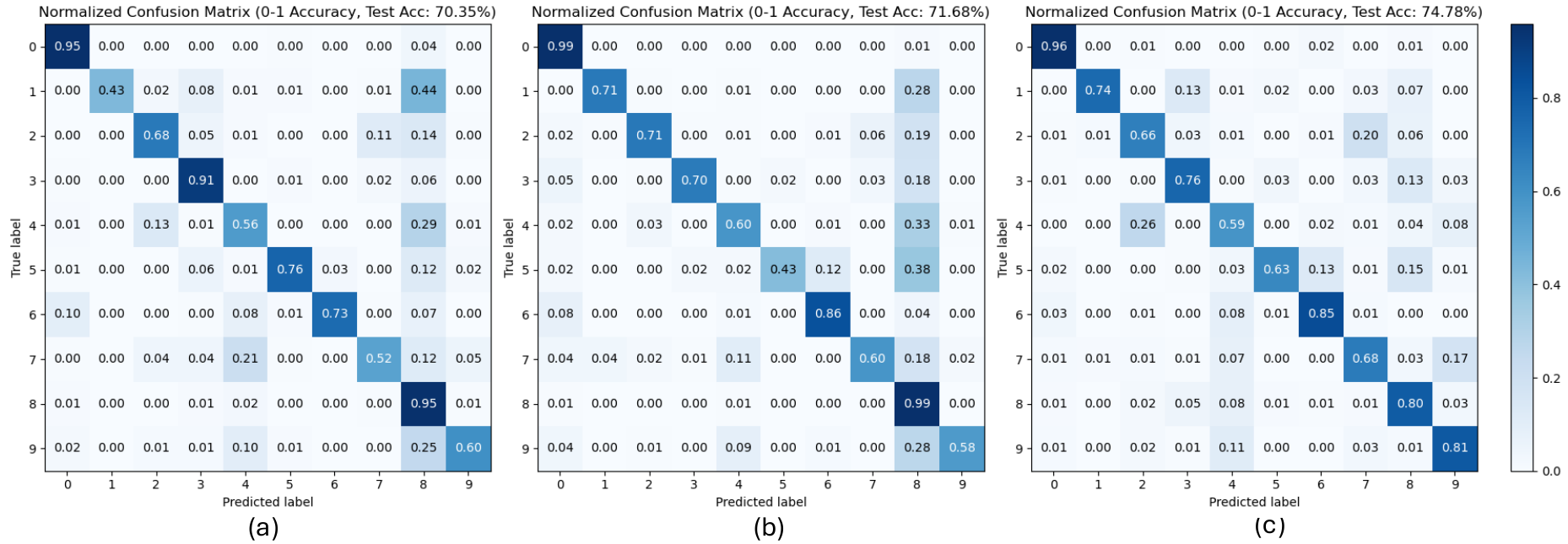}\vspace{-2mm}
  \caption{Confusion matrices where test samples are perturbed with Gaussian noise (mean = 0, standard deviation = 0.25) and rotated by 30 degrees: (a) M1, (b) M2, (c) M3.}\vspace{-3mm}
  \label{fig:con}
\end{figure*}

\begin{table}[t]
\caption{Performance under combined noise and rotation (20°).}\vspace{-2mm}
\label{tb2}
\centering
\small
\begin{tabular}{|c|c|c|c|c|c|}
\hline
\textbf{Method} & 
\shortstack{\textbf{Std=}\\\textbf{0.1+20\textdegree}} & 
\shortstack{\textbf{Std=}\\\textbf{0.2+20\textdegree}} & 
\shortstack{\textbf{Std=}\\\textbf{0.3+20\textdegree}} & 
\shortstack{\textbf{Std=}\\\textbf{0.4+20\textdegree}} & 
\shortstack{\textbf{Std=}\\\textbf{0.5+20\textdegree}} \\
\hline
M1 & 95.24\% & 91.58\% & 79.52\% & 36.03\% & 31.14\% \\
M2 & 96.81\% & 91.15\% & 60.04\% & 26.21\% & 24.83\% \\
M3 & 91.56\% & 90.74\% & 85.52\% & 76.57\% & 61.03\% \\
M4 & 92.18\% & 91.14\% & 83.71\% & 66.20\% & 58.35\% \\
\hline
\end{tabular}
\end{table}

From Tables \ref{tb1}, \ref{tb2}, and \ref{tab:fmnist}, we observe that M2 offers only marginal improvements over M1 under mild Gaussian noise or small geometric perturbations. However, as the severity of corruption increases—such as higher noise levels or more significant rotations—both M1 and M2 experience a steep decline in accuracy. This sharp degradation highlights the limitations of heuristic regularization methods like label smoothing, which apply uniform adjustments regardless of input complexity or ambiguity, and therefore struggle to generalize under distributional shifts.

In contrast, our proposed approaches, M3 and M4, exhibit markedly stronger robustness, especially under challenging conditions. M3, which employs quantum-derived probabilistic labels across all training samples, maintains high and consistent accuracy even as corruption increases. This demonstrates the strength of probabilistic supervision in capturing nuanced class relationships and reflecting uncertainty, allowing the model to generalize better in the presence of noise and transformations. M4 extends this idea by training only on high-confidence samples (covering approximately 91.5\% of the dataset), effectively filtering out noisy or ambiguous data. This strategy yields further robustness improvements by focusing the learning process on reliable examples, reducing the impact of mislabeled or confusing inputs, and improving training efficiency.

\begin{table}
\centering\caption{Classification accuracy on Fashion MNIST under different levels of Gaussian noise (mean=0.0).}\vspace{-2mm}
\begin{tabular}{|c|c|c|c|c|}
\hline
\small
\textbf{Methods} & \textbf{Original} & \textbf{Std = 0.1} & \textbf{Std = 0.2} & \textbf{Std = 0.3} \\
\hline
M1 & 91.32\% & 84.55\% & 70.88\% & 49.80\% \\
\hline
M2 & 91.94\% & 73.36\% & 65.24\% & 40.09\% \\
\hline
M3 & 88.92\% & 87.53\% & 81.80\% & 70.92\% \\
\hline
M4 & 83.51\%
& 82.97\%
& 81.73\%
& 72.41\%
\\
\hline
\end{tabular}
\label{tab:fmnist}
\end{table}

Notably, these trends hold across both MNIST and the more challenging Fashion-MNIST dataset, which contains greater intra-class variation and less distinctive visual cues. This further validates the effectiveness of our quantum-assisted probabilistic labeling framework in real-world settings where input uncertainty is common. Overall, while M1 and M2 may suffice in clean or slightly perturbed environments, M3 and M4 provide more resilient and uncertainty-aware learning mechanisms, making them better suited for applications such as autonomous driving or medical diagnosis where robustness to data variation is critical.

\subsection{Quantitative Analysis}

\paragraph{Confusion matrix}

We report confusion matrices under challenging conditions in which test samples are perturbed with Gaussian noise (mean = 0, standard deviation = 0.25) and rotated by 30 degrees: (a) M1, (b) M2, and (c) M3, shown in Figure \ref{fig:con}. These matrices provide a comprehensive view of model performance by illustrating how predictions are distributed across all classes, allowing us to assess both overall accuracy and specific misclassification trends. Compared to M1 and M2, our M3 model exhibits more balanced and accurate predictions across classes, reflecting enhanced robustness to noise and geometric distortions. This improvement underscores M3's capacity to effectively manage ambiguous or degraded inputs, an essential capability for real-world applications such as autonomous driving and medical imaging, where uncertainty and variability are common.

\paragraph{Label reliability}

\begin{figure}[h!]
  \centering
  \includegraphics[width=1\linewidth]{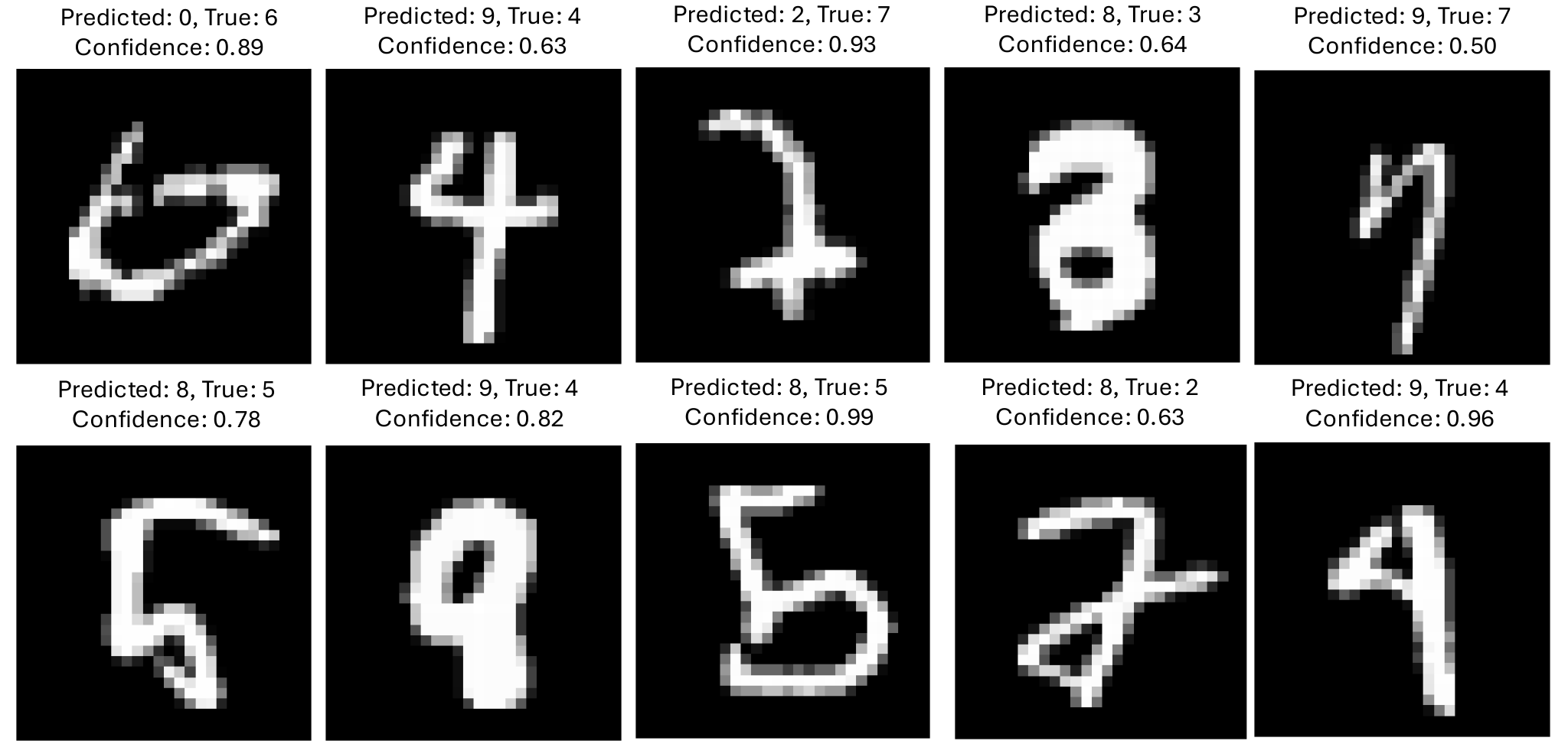}\vspace{-1mm}
  \caption{Samples of Wrong predictions from our M3 with all original test samples, while M1 and M2 generate predictions the same as the given labels with near-1 confidence.}\vspace{-2mm}
  \label{fig:wp}
\end{figure}

We present several challenging test samples for which our M3 model—trained with quantum-derived probabilistic labels—produces uncertain predictions that better capture the inherent ambiguity of the inputs (Figure \ref{fig:wp}). In many of these cases, even human judgment would struggle to confidently assign the correct label. In contrast, the baseline models M1 and M2 generate overconfident predictions with probabilities close to 1, strictly adhering to the given labels and overlooking the uncertainty. This distinction is particularly important in real-world applications such as autonomous driving and medical diagnosis, where recognizing and quantifying uncertainty is essential for preventing critical errors, supporting safer decisions, and fostering trust in AI-assisted systems under ambiguous or noisy conditions.

\section{Conclusion}
We introduced Quantum Probabilistic Label Refining (QPLR), a hybrid framework that leverages quantum non-determinism to generate soft, data-dependent labels. By exploiting the entanglement, superposition, and measurement capabilities of variational quantum circuits (VQCs), QPLR captures input uncertainty and class correlation, offering more expressive and informative labels for classical deep neural networks learning.

Unlike traditional quantum–classical models that embed quantum layers into the training loop, QPLR decouples the quantum component by using it for one-time label generation. This offline quantum labeling allows scalable, architecture-agnostic improvements for classical models. Experiments on MNIST and Fashion-MNIST confirm that QPLR improves robustness under noise and rotation without requiring adversarial methods or architectural changes.

\subsection*{Limitation and Future Work}
Our current pipeline depends on full-state GPU-based simulation, which supports up to 35 qubits but scales poorly due to exponential memory growth. In addition, current quantum hardware remains limited by noise, low fidelity, and costly execution, making large-scale deployment impractical. 

Future work will explore scalable simulation backends, such as tensor network approaches \cite{huang2021efficient}, and integrate noise-aware techniques for real hardware deployment. As quantum processors improve, we expect QPLR to serve as an efficient quantum labeling module—enhancing datasets for broader use in downstream tasks like reinforcement learning and generative modeling.


\appendix

\section*{Appendix: Supplemental Material}

\subsection*{A.1 Computational Cost Comparison}

To quantify how our hybrid quantum–classical framework scales in practice, we measured wall‐clock training time and final accuracy across a range of QPLR configurations, which are shown in Fig. \ref{tab:cost_comparison_ext}.  All experiments were capped at 72h of GPU time on a single NVIDIA A100, and used the same classical pre/post‐network (two 128‐unit dense layers) wrapped around the VQC.  We varied:
\begin{itemize}
  \item \textbf{Qubit count} \(Qubits\) (so that \(2^n\ge K=10\)),
  \item \textbf{Circuit depth} \(Layers\) (number of variational layers),
  \item \textbf{Entanglement pattern} \(Entang.\)(linear, ring, or full CZ),
  \item \textbf{Epochs} \(E\),
  \item \textbf{Batch size} \(B\).
\end{itemize}

\begin{table*}[h]
  \centering
  \small
  \caption{Training time and performance for selected QPLR configurations.  “Time” reports wall‐clock hours and completed epochs.}
  \label{tab:cost_comparison_ext}
  \begin{tabular}{l l c c l c c c l c c}
    \toprule
    Dataset      & Encod. & Qubits & Layers & Entang. & \(E\) & \(B\) & LR     & Time        & Loss    & Acc.\,(\%) \\
    \midrule
    MNIST        & angle  & 10     & 1      & linear  & 5     & 64    & 0.001  & 6h(5/5)   & 0.1162  & 96.34      \\
    MNIST        & angle  & 10     & 1      & linear  & 5     & 128   & 0.001  & 6h(5/5)   & 0.1197  & 96.37      \\
    MNIST        & angle  & 10     & 1      & linear  & 10    & 128   & 0.001  & 13h(10/10)& 0.0742  & 97.14      \\
    MNIST        & angle  & 15     & 1      & linear  & 5     & 64    & 0.001  & 23h(8/15) & 0.1696  & 95.42      \\
    MNIST        & angle  & 15     & 3      & linear  & 15    & 64    & 0.001  & 71h(11/15)& 0.1022  & 97.63      \\
    \midrule
    Fashion      & angle  & 10     & 1      & full    & 25    & 32    & 0.001  & 32h(25/25)& 0.3443  & 88.81      \\
    Fashion      & angle  & 10     & 1      & ring    & 25    & 32    & 0.001  & 32h(25/25)& 0.3443  & 88.81      \\
    Fashion      & angle  & 10     & 1      & linear  & 25    & 32    & 0.001  & 32h(25/25)& 0.3443  & 88.81      \\
    Fashion      & angle  & 20     & 1      & linear    & 25    & 32    & 0.001  & 71h(9/25) & 0.3478  & 88.61     \\
   Fashion      & angle  & 20     & 3     & linear    & 25    & 32    & 0.001  & 71h(5/25) & 0.4391  & 84.67      \\
    \bottomrule
  \end{tabular}
\end{table*}

\paragraph{Analysis and Insights}
\begin{itemize}
  \item \textbf{Qubit scaling:} Increasing \(n\) from 10 to 15 increases simulation time by about eight times (state‐vector size grows exponentially), yet yields no accuracy improvement on MNIST unless paired with deeper circuits and more epochs. In experiments where we combined a 15-qubit VQC with \(L=3\) layers and extended training to 15 epochs, we observed a modest \(\sim\)2\% accuracy gain—likely due to the richer entanglement and longer optimization.
  \item \textbf{Depth scaling:} On Fashion-MNIST, raising the number of variational layers \(L\) from 1 to 3 at \(n=15\) increases wall-clock time by \(\approx2.1\times\) while delivering a \(\sim\)2\% boost in accuracy, confirming nearly linear cost growth with circuit depth.
  \item \textbf{Entanglement pattern:} For shallow circuits, full CZ, ring, and linear topologies exhibit similar runtime and final accuracy. However, as \(L\) increases, the additional two-qubit gates in a fully connected graph will incur progressively higher overhead, potentially widening runtime differences.
  \item \textbf{Dataset complexity:} Fashion-MNIST requires around 25 epochs to converge, substantially extending runtime. Configurations exceeding our 72 h limit were unable to finish, highlighting practical bounds on circuit size, depth, and dataset difficulty.
\end{itemize}

These observations inform the selection of VQC parameters that strike an effective balance between computational cost and model robustness. Moreover, emerging batched GPU simulation techniques promise to accelerate deep-circuit training \cite{jiang2025bqsim, zhao2022q}, enabling more efficient exploration of larger epoch and layer configurations on challenging datasets.

\subsection*{A.2 Quantum Simulation Scalability}

Our QPLR framework relies on classical simulation of variational quantum circuits (VQCs) to generate probabilistic labels.  Below we review (1) basic quantum primitives, (2) NISQ hardware constraints and compilation, (3) GPU‐based state‐vector simulation, and (4) batched‐execution techniques.

\paragraph{Quantum computing primitives.}  
A qubit is a two‐level quantum system in state  
\[
\ket{\psi} = \alpha\ket{0} + \beta\ket{1},\quad \alpha,\beta\in\mathbb{C},\;|\alpha|^2+|\beta|^2=1.
\]  
Single‐qubit gates such as \(R_Y(\theta)\) enact rotations on the Bloch sphere, while two‐qubit gates like CZ generate entanglement.  An \(n\)-qubit VQC applies a sequence of parameterized unitaries  
\[
U(\boldsymbol\theta)=U_L(\theta_L)\cdots U_1(\theta_1)
\]
to \(\ket{0}^{\otimes n}\), yielding
\[
\ket{\psi(\boldsymbol\theta)} = U(\boldsymbol\theta)\,\ket{0}^{\otimes n}.
\]
Measuring in the computational basis produces bitstrings \(y\in\{0,1\}^n\) with Born‐rule probability \(\lvert\braket{y}{\psi(\boldsymbol\theta)}\rvert^2\), and repeated shots estimate the full output distribution \cite{nielsen2010quantum}.

\paragraph{NISQ hardware constraints.}  
Noisy intermediate‐scale quantum (NISQ) devices support on the order of 10–100 qubits without full error correction \cite{preskill2018quantum}.  Qubits—e.g., \ superconducting transmons—are connected in a nearest‐neighbor layout.  To execute a logical circuit, the quantum compiler must (i) map each logical qubit to a physical qubit, (ii) move interacting qubits adjacent to one another by inserting SWAP networks (each SWAP $=$ 3 CNOTs) when necessary, and (iii) schedule gates to fit within limited coherence windows.  During execution, errors arise from imperfect gate and readout operations (quantified by randomized benchmarking), decoherence (\(T_1\) relaxation and \(T_2\) dephasing), and crosstalk between neighboring qubits \cite{qi2023quantum}.  As a result, current real‐hardware experiments incur long queue times, high access costs, and low‐fidelity outcomes, motivating our use of GPU‐based simulation for efficient, repeatable label generation.  However, as quantum hardware rapidly advances, devices are becoming increasingly reliable and cost‐effective. In the near future, it will be practical to run QPLR directly on real quantum processors—harnessing their native parallelism to achieve runtimes and throughput far beyond what is possible with classical simulation.

\paragraph{GPU‐based state‐vector simulation.}  
For qubit counts up to \(n\le20\), state‐vector simulation on a single NVIDIA A100 remains practical and yields exact Born‐rule probabilities (aside from shot noise).  The simulator maintains the full \(2^n\)-dimensional complex amplitude vector in GPU memory, and each single‐ or two‐qubit gate update requires \(O(2^n)\) work.  Consequently, a depth-\(L\) circuit with \(M\) measurement shots incurs  
\[
O\bigl(M \times L \times 2^n\bigr)
\]
compute, which maps to 6–72h of training for \(n\le20\) and \(L\le7\), but becomes intractable beyond \(\sim\!25\) qubits on a single GPU.  Compared to real NISQ hardware, this approach offers (i) noiseless, exact outcomes, (ii) zero queue times and minimal access cost, and (iii) perfect reproducibility—enabling controlled, repeatable label generation for QPLR.

\vspace{1mm}
\begin{table}[t] 
\small
\centering
\caption{Test accuracy of different label refining methods (including BNN and RS) under increasing Gaussian noise.}
\label{tb5}
\vspace{-2mm}
\begin{tabular}{|c|c|c|c|c|c|}
\hline
\textbf{Method} & 
\shortstack{\textbf{Std=}\\\textbf{0.1}} & 
\shortstack{\textbf{Std=}\\\textbf{0.2}} & 
\shortstack{\textbf{Std=}\\\textbf{0.3}} & 
\shortstack{\textbf{Std=}\\\textbf{0.4}} & 
\shortstack{\textbf{Std=}\\\textbf{0.5}} \\
\hline
M1 & 99.14\% & 98.12\% & 85.50\% & 62.59\% & 37.68\% \\
M2 & 99.30\% & 98.60\% & 73.51\% & 51.01\% & 23.32\% \\
\hline
M3 & 97.87\% & 97.35\% & 95.76\% & 81.34\% & 70.13\% \\
M4 & 97.95\% & 97.82\% & 96.05\% & 80.66\% & 52.82\% \\
\hline
BNN & 98.72\% & 95.31\% & 75.04\% & 45.55\% & 46.82\% \\
RS & 98.82\% & 87.02\% & 66.98\% & 58.59\% & 37.63\% \\
\hline
\end{tabular}
\end{table}

\begin{table}[t]
\caption{Label refining methods test performance under combined input noise and rotation (20\textdegree).}
\label{tb6}
\vspace{-2mm}
\centering
\small
\begin{tabular}{|c|c|c|c|c|c|}
\hline
\textbf{Method} & 
\shortstack{\textbf{Std=}\\\textbf{0.1+20\textdegree}} & 
\shortstack{\textbf{Std=}\\\textbf{0.2+20\textdegree}} & 
\shortstack{\textbf{Std=}\\\textbf{0.3+20\textdegree}} & 
\shortstack{\textbf{Std=}\\\textbf{0.4+20\textdegree}} & 
\shortstack{\textbf{Std=}\\\textbf{0.5+20\textdegree}} \\
\hline
M1 & 95.24\% & 91.58\% & 79.52\% & 36.03\% & 31.14\% \\
M2 & 96.81\% & 91.15\% & 60.04\% & 26.21\% & 24.83\% \\
\hline
M3 & 91.56\% & 90.74\% & 85.52\% & 76.57\% & 61.03\% \\
M4 & 92.18\% & 91.14\% & 83.71\% & 66.20\% & 58.35\% \\
\hline
BNN & 94.60\% & 82.21\% & 63.45\% & 41.79\% & 35.54\% \\
RS & 93.94\% & 85.15\% & 66.98\% & 50.07\% & 35.24\% \\
\hline
\end{tabular}
\end{table}

\paragraph{Future work: Batched circuit execution.}  
To break through the serial‐simulation bottleneck, our next phase will integrate batched‐execution techniques that group many circuit evaluations into large tensor operations.  Specifically:
\begin{itemize}
  \item \emph{Shot‐batch fusion:} Combine all \(M\) measurement shots for each input into a single GPU‐resident tensor, eliminating per‐shot kernel launch overhead.
  \item \emph{Gate‐fusion batching:} Identify and merge identical gate operations across different inputs or parameter sets—applying them in bulk to the state vector in one fused kernel.
  \item \emph{Compute–transfer overlap:} Leverage CUDA streams or task graphs to overlap data movement (e.g.\ amplitude fetch/store) with gate computations, ensuring the GPU remains fully utilized.
\end{itemize}
Recent systems like Q-GPU \cite{zhao2022q} and BQSim \cite{jiang2025bqsim} have demonstrated 3–300× speedups on deep circuits by these methods; incorporating them will allow QPLR to explore deeper VQCs (\(L>7\)) and larger epoch regimes on challenging datasets, within practical wall‐clock budgets and without reliance on costly, noisy NISQ devices.  

\subsection*{A.3 Additional Label Refining Baselines}

While our main text focuses on label smoothing as the most direct and widely-used method for converting one-hot labels to soft labels, other label refining techniques—such as Bayesian Neural Networks (BNN) and Randomized Smoothing (RS)—also warrant comparison. Unlike label smoothing, which deterministically smooths label vectors, BNN and RS generate soft labels by aggregating predictions from models under input or weight perturbations, capturing uncertainty in a data-driven way. Although these approaches do not assign soft labels directly from ground-truth as label smoothing does, they represent important alternatives in the broader landscape of label refinement. For completeness, we report their results in the appendix.

\textbf{Experimental Design:} 
To provide a fair and comprehensive comparison, we construct two additional classical baselines:
\textbf{Bayesian Neural Networks (BNN):} A CNN is trained with MC Dropout enabled. For each training sample, we perform multiple stochastic forward passes with dropout activated, and average the output probabilities to produce a soft label. This reflects the model's epistemic uncertainty.
\textbf{Randomized Smoothing (RS):} A CNN is trained while injecting Gaussian noise (std=0.05) to each input image. At inference, we aggregate softmax predictions over multiple noisy samples, capturing the local input-space uncertainty.

The resulting soft labels from BNN and RS are then used to train new downstream CNNs, with performance evaluated under various test-time noise and rotations, as shown in Tables~\ref{tb5} and~\ref{tb6}.

\textbf{Baseline Key:} 
In the tables, \textbf{M1} refers to hard label (vanilla) baseline; \textbf{M2} is label smoothing; \textbf{M3} is our QPLR; \textbf{M4} is a variant of QPLR; \textbf{BNN} and \textbf{RS} are described above. Note that M1, M2, and their results in Table~\ref{tb5} and Table~\ref{tb6} are consistent with the main text.

\textbf{Analysis:}
The results show that while BNN, RS, and label smoothing (M2) all yield competitive performance at low noise, their accuracy drops precipitously under moderate to severe noise or combined perturbations---with performance falling well below 50\% at the highest noise levels. In contrast, QPLR (M3, M4) maintains markedly higher accuracy, especially in the presence of strong noise and rotations. This suggests that soft labels generated by BNN and RS, which are tied to model uncertainty or input noise, fail to provide the same level of robustness as QPLR’s quantum-derived labels. We believe this is because QPLR’s label refinement mechanism captures high-order correlations in the input features, not just marginal uncertainties, resulting in labels that are more robust to downstream data shifts and adversarial conditions.




\end{document}